# Photoelectron emission from plasmonic nanoparticles: Comparison between surface and volume photoelectric effects

Alexander V. Uskov,[1, 2, 3, 4 a)] Igor E. Protsenko,[1, 2] Renat Sh. Ikhsanov,[5]
Viktoriia E. Babicheva,[6] Sergei V. Zhukovsky,[6] Andrey V. Lavrinenko,[6] Eoin P. O'Reilly,[7]
and Hongxing Xu[3, 4]

[1)]*P. N. Lebedev Physical Institute, Russian Academy of Sciences, Leninskiy Pr. 53, 119333 Moscow, Russia*
[2)]*Advanced Energy Technologies Ltd, Skolkovo, Novaya Ul. 100, 143025, Moscow Region, Russia*
[3)]*School of Physics & Technology, Wuhan University, Wuhan, 430072, P. R. China*
[4)]*Institute of Physics, Chinese of Academy of Sciences, P. O. Box 603-146, Beijing, 100190, P. R. China*
[5)]*Research Institute of Scientific Instruments, State Nuclear Energy Corporation "Rosatom", Moscow, Russia*
[6)]*DTU Fotonik, Technical University of Denmark, Ørsteds Plads 343, DK-2800 Kgs. Lyngby, Denmark*
[7)]*Tyndall National Institute, Cork, Ireland*

We study emission of photoelectrons from plasmonic nanoparticles into surrounding matrix. We consider two mechanisms of the photoelectric effect from nanoparticles – surface and volume ones, and use models of these two effects which allow us to obtain analytical results for the photoelectron emission rates from nanoparticle. Calculations have been done for the step potential at surface of spherical nanoparticle, and the simple model for the hot electron cooling have been used. We highlight the effect of the discontinuity of the dielectric permittivity at the nanoparticle boundary in the surface mechanism, which leads to substantial (by ~5 times) increase of photoelectron emission rate from nanoparticle compared to the case when such discontinuity is absent. For plasmonic nanoparticle, a comparison of two mechanisms of the photoeffect was done for the first time and showed that surface photoeffect, at least, does not concede the volume one, which agrees with results for the flat metal surface first formulated by Tamm and Schubin in their pioneering development of quantum-mechanical theory of photoeffect in 1931. In accordance with our calculations, this predominance of the surface effect is a result of effective cooling of hot carriers, during their propagation from volume of the nanoparticle to its surface in the scenario of the volume mechanism. Taking into account both mechanisms is essential in development of devices based on the photoelectric effect and in usage of hot electrons from plasmonic nanoantenna.

## I. INTRODUCTION

A recent publication by Chalabi and Brongersma [1] has been entitled "Harvest season for hot electrons", and this title excellently illustrates a boom of interest to generation of hot photoelectrons in plasmonic nanostructures occurring at present. Indeed, enhanced photoelectron emission from single plasmonic nanoantennas and from ensembles of such nanoantennas are under intensive studies for application in Schottky barrier photodetectors in order to reach higher device sensitivity [2-14], in solar cells with the goal to enhance the their photovoltaic efficiency by harvesting solar photons below semiconductor bandgap energy [15-17, 2, 4, 6, 11, 14, 18-21], in (nano-)photoelectrochemistry and (nano-)photochemistry [2, 4, 22-30] including, in particular, water splitting [25-26, 28-30], for realization of new photoconductive plasmonic metamaterials [6], in molecular electronics [31] and so – in all areas of science and technology where generation of hot photoelectrons with their subsequent utilization plays principal role. As well, emission of hot electrons can enhance characteristics of solar concentrator systems [32]. It is also worth to note developments and proposals based on the use of emission of photoelectrons from plasmonic nanoantennas (nanotips, first of all) into vacuum – novel nanometer-sized femtosecond electron sources [33], femtosecond photoelectron emission spectroscopy [34], attosecond nanoplasmonic-field microscope [35].

Obviously, understanding physical mechanisms that result in emission of photoelectrons from plasmonic nanoparticles and nanostructures is essential for development of devices based on this phenomenon. The research on this topic dates back to the pioneering work on quantum-mechanical theory of photoelectric effect from metals written in 1931 by Tamm and Schubin [36], who introduced and described two mechanisms of the effect – see Fig.1:

(A) *surface* mechanism (or the *surface photoelectric effect*, see Fig.1a), in which an electron absorbs a photon during its collision with metal surface (boundary), and if the energy received by the electron is sufficient to overcome the potential barrier at the boundary (Schottky barrier if the metal is in contact with semiconductor), the electron is emitted from the metal into the matrix neighboring with the metal (semiconductor, for instance) during this inelastic collision with metal surface. In this case, the electron also can be reflected back to metal after photon absorption during the collision [37]. In the surface mechanism of photoeffect, the rate of photoelectron emission from metal is proportional to the square of the electromagnetic field component, normal to the metal surface [36] – see below.

---
[a)] Electronic mail: alexusk@lebedev.ru

(B) *volume* (or *bulk*) mechanism (or the *volume photoelectric effect*, see Fig.1b), which consists of *three phases* (see an comprehensive review [38] primarily devoted to this bulk mechanism):

(1) an electron absorbs a photon inside the metal during its collision with impurity, phonon, lattice defect, etc [39] or due to its coupling to periodic lattice potential [36, 40], and becomes "hot";

(2) then the electron moves to the boundary of metal (this "electron transport" phase is absent in the surface mechanism), colliding with phonons and cold electrons and losing energy in the process;

(3) if the electron reaches the metal surface with energy that is still sufficient to overcome the potential barrier at the boundary of metal, the electron may be emitted into the matrix (semiconductor) surrounding the metal.

Obviously, the photoelectron emission rate from metal in the scenario of bulk photoeffect is proportional to the light absorption coefficient of bulk metal; it depends on the energy distribution of hot electrons after their generation and on the cooling rate of electrons during their motion to nanoparticle boundary.

Having identified and compared these two mechanisms, Tamm and Schubin in [36] considered the mechanism (A) as dominating in the visible and IR ranges. Nevertheless, during several decades after publication [36], researchers returned to the discussion on the above mechanisms of photoelectric effect from metal (see [38, 41-50] and references therein), and in particular, to the arguments which of the two mechanisms is more important. The main argument against the surface mechanism (A) in those discussions was that the component of field normal to the surface (to the square of which the photoelectron emission rate in the surface mechanism is proportional) is absent if light is incident *normally* on a *flat* metal surface, as was relevant in many practical cases. However, in 60[th]−70[th] of the last century it was understood clearly (see [46-50] and references therein) that roughness of metal surface can lead to the appearance of the normal component to metal surface (in particular, due to conversion of incident plane wave into surface plasmonic wave [49]), so that the mechanism (A) can be essential even in macroscopically flat structures with normal incidence of light.

Nevertheless, the question on which of the two mechanisms of photoelectron emission is dominant has been left open until the present time, with various groups adopting different approaches. Several years ago, Berini with his coauthors laid the volume mechanism (B) as the basis of their consideration of thin-film Schottky barrier photodetectors [7-9], and very recently Halas and Nordlander with their colleagues [3, 12-13] used the description of the volume mechanism, given in [7] to analyze emission of photoelectrons from plasmonic nanoantennas. On the other hand, theory of emission of photoelectrons from metallic nanoparticles developed in [6] and then used in calculations in [11, 14] is based namely on the surface mechanism (A). In the paper [51] Govorov and coauthors develop an approach to the theory of photoelectron emission from plasmonic nanoparticle starting from the quantum microscopic description of non-equilibrium carrier population in a localized plasmon wave. As nanostructures of more and more intricate shapes are introduced, the discussion on proper description of photoelectron emission from metals in general and from metal nanostructures in particular is again becoming of current importance.

In this paper, we compare the surface and volume mechanisms of photoelectron emission from plasmonic nanoparticles. Using spherical particles as a simple and analytically tractable example, we derive comparable metrics for each of the two mechanisms. Comparing these metrics, we conclude that the volume mechanism could only prevail if hot electrons were able to reach the nanoparticle surface without energy loss. In realistic cases, the "cooling" processes during the "hot" electron transport lead to the prevalence of surface photoelectric effect.

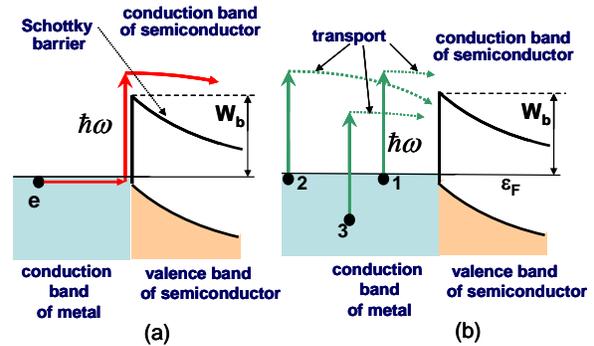

Fig. 1. Illustration of two mechanisms of photoelectric effect. (a) Surface effect: electron collides with the Schottky barrier, absorbs the photon energy $\hbar\omega$, and leaves the metal. (b) Volume effect: electron 1 receives the energy $\hbar\omega$, moves to the Schottky barrier, and overcomes it, leaving metal; electrons 2 and 3 have not sufficient energy when they reach the barrier and remain in metal. $W_b$ is the work function for the metal to semiconductor, $\varepsilon_F$ is the Fermi level.

The paper is organized as follows. In Sect. II we provide a detailed account on the theory of photoelectron emission from plasmonic nanoparticles. In particular, in Sect. II-A, the problem is formulated, and *the concept of the cross-section of photoelectron emission* from a nanoparticle is introduced. In Sect. II-B, formulas to calculate the photoelectron emission rate from nanoparticle and the photoelectron emission cross-section of spherical nanoparticle for surface photoelectric effect are presented based on work [6] for simple model of the step potential at the nanoparticle boundary. In Sect. II-C, a model to calculate the internal quantum efficiency for volume mechanism of photoeffect is presented, along with deriving the expression for the photoelectron emission cross-section for the volume mechanism. In Sect. III, numerical results are presented to compare the two mechanisms, and the basic assumptions of calculations are discussed in details. Finally, Sect. IV formulates the conclusions.

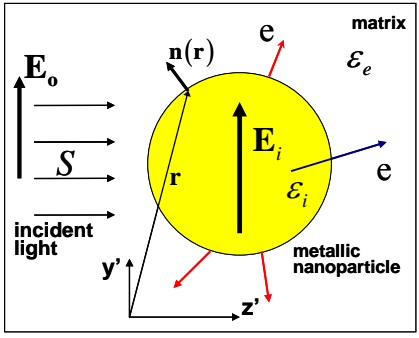

Fig. 2. Schematic illustration of spherical metallic nanoparticle (with permittivity $\varepsilon_i$) imbedded in a dielectric (or semiconductor) matrix (with permittivity $\varepsilon_e$). The incident plane wave with electrical field $\mathbf{E}_o$ causes an electric field $\mathbf{E}_i$ inside the nanoparticle. red arrows illustrate the surface mechanism of photoelectric effect, and blue arrow shows the volume (bulk) mechanism of photoeffect.

## II. Theory of photoelectric effect

### A. Formulation of the problem

The following problem is under consideration – see Fig.2. A plane light wave of frequency $\omega$ and with intensity $S$ is propagating along the $z'$-axis in a background matrix (dielectric or semiconductor) with relative permittivity $\varepsilon_e$. The amplitude $\mathbf{E}_o$ of the electric field of the light is polarized along the $y'$ direction. The wave is incident on an imbedded metal nanoparticle with relative permittivity $\varepsilon_i(\omega)$. For simplicity, we consider *spherical* nanoparticle with the radius $a$ so that in quasistatic approximation [52] the field $\mathbf{E}_i$ inside nanoparticle is homogeneous, parallel to $\mathbf{E}_o$ and can be expressed as [52]

$$\mathbf{E}_i = \frac{3\varepsilon_e}{\varepsilon_i(\omega) + 2\varepsilon_e}\mathbf{E}_o \equiv F \cdot \mathbf{E}_o, \qquad (1)$$

When the light frequency $\omega$ approaches the frequency $\omega_{lpr}$ which satisfies the Fröhlich condition [52]

$$\mathrm{Re}\left[\varepsilon_i(\omega_{lpr})\right] + 2\varepsilon_e = 0, \qquad (1a)$$

as one can see from (1), the resonance enhancement of the field takes places (the localized plasmonic resonance). Note that the homogeneous field assumption may be violated in presence of a strong plasmonic resonance and/or in nanoparticles of more complex shapes (e.g., nanoantennas), where strong field localization effects can lead to field inhomogeneity inside the nanoparticle. However, the presented formalism is straightforwardly generalized to the case of inhomogeneous field, albeit at the cost of no longer being analytically tractable.

The electrons of metal absorb photons with the energy $\hbar\omega$ (A) during their collisions with the nanoparticle surface (*surface photon absorption*) and (B) inside the nanoparticle (*volume photon absorption*) [36-37], and can leave the nanoparticle in either case. Our goal is to calculate the rates of the photoelectron emission from the nanoparticle due to the surface and the volume absorption of photons and compare them. In the next two subsections, we present separately the calculations for surface and volume (bulk) photoelectric effects from metal nanoparticle.

Ability of plasmonic nanoparticles to create emission of photoelectrons can be characterized by the photoelectron emission cross-section of nanoparticle [6]. Namely, the photoelectron emission cross-section is

$$\sigma_{em} = R_{em}/(S/\hbar\omega), \qquad (2)$$

where $R_{em}$ is the rate of emission of photoelectrons from nanoparticle in $(1/\mathrm{s})$, $S/\hbar\omega$ is the photon flux [in $1/(\mathrm{m}^2 \cdot \mathrm{sec})$] incident on the nanoparticle. Below we calculate $R_{em}$ and $\sigma_{em}$ both for surface and volume photoelectric effect.

### B. Theory of surface photoelectric effect

In this subsection, we briefly introduce the theory of surface mechanism of photoelectron emission following to [6] where the theory is presented in more details. If the de Broglie electron wavelength $\lambdabar$ in the metal is much smaller than the characteristic nanoparticle size $L_{nano}$, $\lambdabar \ll L_{nano}$ (one should note that in silver and gold $\lambdabar \approx 0.5\mathrm{nm}$), we can safely neglect quantum-confinement effects in the metal. In other words, the electron gas is uniformly distributed with an equilibrium density given by that of the bulk metal. Furthermore, we can calculate the rate $u(\mathbf{r})$ of electron emission per unit area of nanoparticle surface $[1/(\mathrm{s}\times\mathrm{m}^2)]$, by considering the nanoparticle surface at the coordinate $\mathbf{r}$ (see Fig.2) as being *flat*, and by using the theory of photoelectron emission due to collisions of metal electrons with a *flat* boundary. Within this approximation, the rate $u(\mathbf{r})$ is proportional to the square of the *normal* component $E_i^{(n)}(\mathbf{r}) = \mathbf{n}(\mathbf{r})\cdot\mathbf{E}_i$ of the field $\mathbf{E}_i$ [6, 36, 41-43]:

$$u(\mathbf{r}) = C_{em}^{surface} \cdot \left|E_i^{(n)}(\mathbf{r})\right|^2, \qquad (3)$$

where $\mathbf{n}(\mathbf{r})$ is the unit vector normal to nanoparticle surface. The coefficient $C_{em}^{surface}$ is calculated quantum-mechanically (see below) and depends, in particular, on the electron density in the metal, on the photon energy $\hbar\omega$, on the potential barrier for electrons at the nanoparticle boundary, and on any discontinuities in the permittivity and the electron mass at the interface between the metal and the surrounding matrix. Correspondingly, the photoelectron emission rate due to electron collisions with the total nanoparticle surface is

$$R_{emission}^{surface} = \int_{surface} ds\, u(\mathbf{r}) = C_{em}^{surface} \int_{surface} ds\, \left|E_i^{(n)}(\mathbf{r})\right|^2, \quad (4)$$

where the integral extends over the entire nanoparticle surface. Since the field inside spherical nanoparticle is homogeneous, we get easily

$$R_{em}^{surface} = C_{em}^{surface} \cdot A_{nano} \left|\mathbf{E}_i\right|^2 / 3, \quad (5)$$

where $A_{nano} = 4\pi a^2$ is the area of nanoparticle surface.

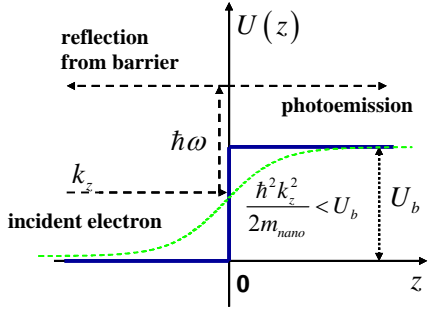

Fig. 3. Schematic illustration of inelastic scattering of an electron at a metal boundary in the presence of an optical field. The potential energy profile $U(z)$ is plotted along the direction $z$ normal to the metal boundary. An electron incident on the boundary (wave vector $k_z$) scatters in-elastically, by absorption of a photon (energy $\hbar\omega$). In the collision with the boundary, the electron can be partly back-reflected into the metal or be forward scattered into the dielectric matrix (emission of photoelectrons). Blue line is the step potential with the height $U_b$, and green curve illustrates an example of gradually changing potential

The coefficient $C_{em}^{surface}$ in (2) can be found by solving the quantum-mechanical problem for the collision of a single electron with a metal boundary, and then subsequently summing over all metal electrons undergoing such collisions with the surface. Fig. 3 illustrates the problem. The metal boundary is modeled by the 1D potential barrier $U(z)$ where the axis $z$ is normal to the boundary, and in this section, we perform calculation for an abruptly changing (at $z=0$) potential (*the step potential*) with a step of height $U_b$, as indicated in Fig. 3. The potential step $U_b$ can be written as $U_b = \varepsilon_F + W_b$ where $\varepsilon_F$ is Fermi energy in metal, $W_b$ is the work function for the metal bordering a given surrounding medium. The electron masses in metal $m_i$ and in surrounding medium (barrier) $m_e$ can be different from each other, in general. We consider an electron plane wave in the metal incident on the metal boundary with wave vector $\mathbf{k}_i = (k_{ix}, k_{iy}, k_{iz})$. In the absence of an electromagnetic field, the electron scattering from the barrier is elastic and furthermore the parallel wave vector component is conserved (since the surface is assumed locally flat). On the other hand, the electron may scatter inelastically in the presence of an electromagnetic field, i.e. by absorbing a photon with energy $\hbar\omega$. While the parallel momentum of the electron is still conserved (neglecting the vanishing momentum of the photon itself), it may either scatter back into the metal or out into the surrounding matrix – see Fig.3. We denote the corresponding probabilities by $p_{in}$ and $p_{out}$, respectively. Both probabilities $p_{in}$ and $p_{out}$ are proportional to the square of the *normal* component $E_i^{(n)}(\mathbf{r}) = \mathbf{n}(\mathbf{r}) \cdot \mathbf{E}_i$ of the field $\mathbf{E}_i$ in the metal [6, 36, 41-45]:

$$p_{in} = c_{in} \cdot \left|E_i^{(n)}(\mathbf{r})\right|^2, \quad (6)$$

$$p_{out} = c_{out} \cdot \left|E_i^{(n)}(\mathbf{r})\right|^2, \quad (7)$$

Note that the normal component $E_i^{(n)}$ inside metal is related to the normal component $E_e^{(n)}$ in surrounding medium by

$$\varepsilon_i E_i^{(n)} = \varepsilon_e E_e^{(n)}, \quad (8)$$

Below we are concentrating on the calculation of the photoelectron emission probability $p_{out}$. Obviously, with the step potential as in Fig.3, photoelectron emission from the metal can occur (i.e., $c_{out} > 0$) only if the electron gains sufficient energy to overcome the barrier, i.e. only if $\hbar^2 k_{iz}^2/(2m_i) + \hbar\omega > U_b$.

Although the probability $p_{out}$ can be calculated with various quantum-mechanical methods, in [6] it is found through direct solution of the Schrödinger equation for an electron in the presence of the field using the perturbation theory (see also very detailed description in [41]). In this solution, the electron wave function in barrier far from the boundary ($z \to \infty$) contains the component

$$C_+(\infty) \cdot \exp\left[-i\frac{\varepsilon(\mathbf{k}_i) + \hbar\omega}{\hbar}t + i\left(k_{ix}x + k_{iy}y + k_{ez}^+ z\right)\right], \quad (9)$$

describing electron of the initial energy in metal

$$\varepsilon(\mathbf{k}_i) = \hbar^2\left(k_{ix}^2 + k_{iy}^2 + k_{iz}^2\right)/(2m_i), \quad (10)$$

which absorbed the photon energy $\hbar\omega$ and left metal for the surrounding medium. The $z$-component $k_{ez}^+$ of the electron wave vector after the electron is emitted outside the metal after absorption of the photon is determined from the energy conservation law:

$$\frac{\hbar^2\left(k_{ix}^2 + k_{iy}^2 + k_{iz}^2\right)}{2m_i} + \hbar\omega = \frac{\hbar^2\left[k_{ix}^2 + k_{iy}^2 + \left(k_{ez}^+\right)^2\right]}{2m_e} + U_b, \quad (11)$$

The amplitude $C_+(\infty)$ is proportional to the normal component $E_i^{(n)}$ of the field in metal and can be written as

$$C_+(\infty) = \left(b_V U_b + b_m \Delta m + b_\varepsilon \Delta\varepsilon\right) \cdot E_i^{(n)}, \quad (12)$$

where $\Delta m = m_e - m_i$ and $\Delta\varepsilon = \varepsilon_e - \varepsilon_i$; $b_V$, $b_m$, and $b_\varepsilon$ are some coefficients, in general – complex numbers. The emission probability $p_{out}$ is expressed through the amplitude $C_+(\infty)$ as

$$p_{out} = \frac{\text{Re}\left[k_{ez}^+\right]}{k_{iz}}\left|C_+(\infty)\right|^2, \qquad (13)$$

(we assume *unit* probability amplitude of the electron plane wave wavefunction incident from the metal to the barrier). Correspondingly,

$$c_{out} = \frac{k_{ez}^+}{k_{iz}}\left|b_V U_b + b_m \Delta m + b_\varepsilon \Delta\varepsilon\right|^2, \qquad (14)$$

Eq. (14) demonstrates clearly that photon absorption by electron with emission from metal takes place (a) due to the jump $U_b$ of the potential, (b) due to the discontinuity $\Delta m$ of the electron mass, and (c) due to the discontinuity $\Delta\varepsilon$ of the dielectric constant at the nanoparticle surface. The photon absorption due to nonzero $\Delta\varepsilon$ can be considered as an effect which is inverse to the *transit radiation* effect, when electron crosses boundary between two mediums with different dielectric constants and emits light [53]. Below, we show that nonzero $\Delta\varepsilon$ substantially increases the surface photoelectric effect.

Below for simplicity we assume that $\Delta m = 0$, so that $m_e = m_i \equiv m$. In this case, calculating the coefficients $b_V$ and $b_\varepsilon$ as in [6], we are coming to the formula

$$c_{out} = \frac{8e^2 U_b}{m\hbar^2\omega^4}\cdot\text{Re}\left[\sqrt{\left(X-1+\frac{\hbar\omega}{U_b}\right)}\right]\cdot\frac{G(X)}{\sqrt{X}}\cdot\left|K_{\Delta\varepsilon}(X)\right|^2, (15)$$

where $e$ is the electron charge, $X = \varepsilon_{iz}(k_{iz})/U_b$ with $\varepsilon_{iz}(k_{iz}) = \hbar^2 k_{iz}^2/(2m)$ so that the coefficient $c_{out}$ depends only on the *z*-component $k_{iz}$ of the initial vector $\mathbf{k}_i$, normal to the boundary: $c_{out} = c_{out}\left[\varepsilon_{iz}(k_{iz})\right]$;

$$G(X) = X \cdot \frac{\left|\sqrt{X}-\sqrt{X-1}\right|^2}{\left|\sqrt{X+\hbar\omega/U_b}+\sqrt{X+\hbar\omega/U_b-1}\right|^2}, \qquad (16)$$

and the coefficient

$$K_{\Delta\varepsilon} = \frac{1}{2}\left[\left(\frac{\varepsilon_i}{\varepsilon_e}+1\right)-\left(\frac{\varepsilon_i}{\varepsilon_e}-1\right)\cdot\left(\sqrt{X+\frac{\hbar\omega}{U_b}}+\sqrt{X-1}\right)^2\right], \qquad (17)$$

describes the effect of the discontinuity $\Delta\varepsilon$ on the photon absorption and photoelectron emission – if $\varepsilon_e = \varepsilon_i$, we have $K_{\Delta\varepsilon} \equiv 1$.

Summing over all electrons in the metal that collide with the surface in the metal, one can obtain the coefficient in Eq.(3) as

$$C_{em}^{surface} = \int_{k_{iz}>o} \frac{2 d\mathbf{k}_i}{(2\pi)^3} f_F(\mathbf{k}_i)\cdot v_{iz}\cdot c_{out}, \qquad (18)$$

where $f_F(\mathbf{k}_i) = \left\{1+\exp\left[(\varepsilon(\mathbf{k}_i)-\varepsilon_F)/k_B T_e\right]\right\}^{-1}$ is the Fermi-Dirac equilibrium distribution function for electrons in the metal, $T_e$ is the electron temperature; $v_{iz} = \hbar k_{iz}/m$ is the electron velocity component normal to the metal boundary. Since the coefficient $c_{out}$ depends only on $k_{iz}$ (not on $k_{ix}$ and $k_{iy}$), the 3D integrals in (18) can be easily converted into 1D-integrals over $k_{iz}$.

Let us introduce the dimensionless coefficient $\eta_o$ related to $C_{em}^{surface}$ as [6]

$$\eta_o = \frac{\hbar\omega}{2\varepsilon_o c}\cdot C_{em}^{surface}, \qquad (19)$$

through which the *external quantum efficiency* (the quantum yield) for devices based on photoelectron emission from ensembles of nanoantennas can be expressed [11, 14]. The parameter $\eta_o$ itself can be interpreted as the external quantum efficiency of a device in which the incident light with the intensity $S_{vac} = 2\varepsilon_o c|E_{vac}|^2$ in vacuum ($E_{vac}$ is the electric field of light) creates the normal component $E_i^{(n)} = E_{vac}$. Correspondingly, from Eqs (15), (18) and (19) we have for $T_e = 0$

$$\eta_o = \frac{8}{\pi}\alpha_{f-s}\cdot\left(\frac{U_b}{\hbar\omega}\right)^3\cdot$$
$$\cdot\int_{1-\hbar\omega/U_b}^{\varepsilon_F/U_b} dX\,\text{Re}\left[\sqrt{X-1+\frac{\hbar\omega}{U_b}}\right]\frac{G(X)}{\sqrt{X}}\left|K_{\Delta\varepsilon}(X)\right|^2\left(\frac{\varepsilon_F}{U_b}-X\right), \qquad (20)$$

where $\alpha_{f-s} = e^2/(4\pi\varepsilon_o\hbar c) = 0.007297 \sim 1/137$ is the fine-structure constant, and we assume that $\hbar\omega < \varepsilon_F$. Blue curve in Fig. 4 illustrates the dependence of $\eta_o$ on the photon energy $\hbar\omega$. In calculations, we assumed that gold nanoparticle is surrounded with a medium with $\varepsilon_e = 13$ (like GaAs). For gold, we used the dielectric constant $\varepsilon_i(\omega)$ from [54]. Correspondingly, [see the condition (1a)], the plasmonic resonance in spherical nanoparticles occurs at $\hbar\omega_{lpr} = 1.48\,\text{eV}$. We also used the values $\varepsilon_F = 5.51\,\text{eV}$ and $W_b = 0.8\,\text{eV}$. One can see that the parameter $\eta_o$ changes from zero at the threshold ($\hbar\omega = 0.8\,\text{eV}$) to $\sim 0.002$ at $\hbar\omega = 1.4\,\text{eV}$. Red curve illustrates Eq.(21) obtained from (20) by approximate integration:

$$\eta_o \approx \frac{32\alpha_{s-f}}{15\pi}\left(\frac{U_b}{\hbar\omega}\right)^3\sqrt{\frac{1}{\bar{X}}}\,G(\bar{X})\left|K_{dis}(\bar{X})\right|^2\left(\frac{\hbar\omega-W_b}{U_b}\right)^{5/2},$$
$$(21)$$

where $\bar{X} = 0.5\cdot\left[1+(\varepsilon_F-\hbar\omega)/U_b\right]$. Formula (21) shows clearly that near to the threshold, when the photon energy $\hbar\omega$ approaches the work function $W_b$,

$$\eta_o \propto (\hbar\omega-W_b)^{5/2} \qquad (22)$$

Eq.(21) is different from the *parabolic* Fowler 's law where $\eta_o \propto (\hbar\omega-W_b)^2$ [55], and takes place, in general, when the potential $U(z)$ at the metal boundary changes rather sharply than gradually. In contrast, the parabolic

Fowler's law works for gradually changing potentials (see as example green curve in Fig.3), see [43-45], for instance.

Green curve in Fig.4 shows the parameter $\eta_o$ when $\Delta\varepsilon = 0$ [i.e., $K_{\Delta\varepsilon}(X) \equiv 1$ in Eq.(20)]. Comparison of blue and green curves demonstrates that nonzero discontinuity $\Delta\varepsilon$ of the dielectric constant $\varepsilon$ at the boundary between metal and surrounding medium substantially (by 3-10 times) increases the surface photoelectron emission parameter $\eta_o$.

Blue and green curves in Fig. 4 show also that the initial fast growth of $\eta_o$ saturates with increasing photon energy $\hbar\omega$, and after this saturation the parameter $\eta_o$ decreases (not shown). This behavior is due to the strong suppression of the interaction of electron with electromagnetic field with increasing photon energies $\hbar\omega$ – see the $1/\omega^4$-dependence in Eq.(15). This circumstance, in particular, leads to a diminished role of surface mechanism compared to the bulk mechanism with increasing photon energy $\hbar\omega$ – see Sect. III.

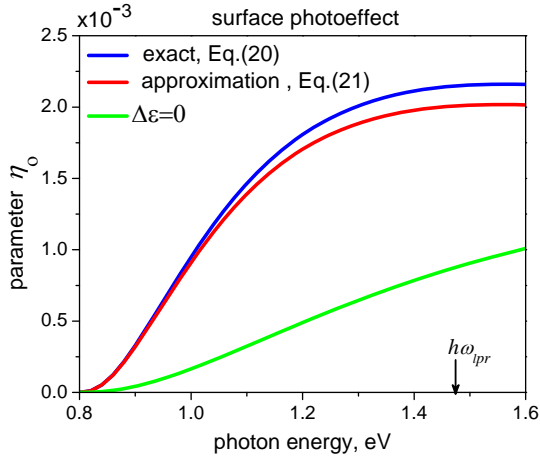

Fig. 4. Spectral dependence of the photoelectron emission parameter $\eta_o = \eta_o(\hbar\omega)$. Blue curve is calculated from Eq.(20), red one is the approximation (21). The green curve is obtained in assumption $\Delta\varepsilon = 0$ [i.e., $K_{\Delta\varepsilon}(X) \equiv 1$ in Eq.(20)]. $\hbar\omega_{lpr}$ is the photon energy when the plasmonic resonance occurs in the nanoparticle.

Following the definition (2) and using (5) and (19), we obtain a formula for the photoelectron emission cross-section due to the surface photoelectric effect:

$$\sigma_{em}^{surface} = \frac{4\pi\eta_o}{3\varepsilon_e^{1/2}}|F|^2 \cdot a^2, \quad (23)$$

where we have used the expression $S = 2\varepsilon_o\varepsilon_e^{1/2}c|\mathbf{E}_o|^2$. Enhancement of the photoelectron emission cross-section due to plasmonic nanoantenna effect is presented in Eq. (23) by the factor $|F|^2$ (see also Eq. (1)). Examples of calculation of $\sigma_{em}^{surface}$ can be found in [6].

*C. Calculation of volume photoelectric effect*

In our modeling of volume photoelectric effect from metal nanoparticle we are closely following the approach given in the paper by Chen and Bates [56] (see also the three-step description of the volume mechanism in [38, 57], and also references therein and in [56]). The power absorbed inside the nanoparticle is given by [52]

$$P = 2\omega\varepsilon_o\varepsilon_i'' \cdot \int_{volume} d\mathbf{r}\cdot|\mathbf{E}_i|^2 = 2\omega\varepsilon_o\varepsilon_i''\cdot|\mathbf{E}_i|^2 V_{nano}, \quad (24)$$

where $V_{nano} = 4\pi a^3/3$ is the nanoparticle volume. Correspondingly, the photon absorption rate (in $1/s$) in the whole nanoparticle is

$$R_{abs}^{volume} \equiv P/\hbar\omega = 2\hbar^{-1}\varepsilon_o\varepsilon_i''\cdot|\mathbf{E}_i|^2 V_{nano} = r_{abs}^{vol}\cdot V_{nano}, \quad (25)$$

where

$$r_{abs}^{vol} = 2\hbar^{-1}\varepsilon_o\varepsilon_i''\cdot|\mathbf{E}_i|^2, \quad (26)$$

is the *volume density* of photon absorption rate [$1/(s\cdot m^3)$] in the nanoparticle. We assume that electrons in metal before their excitation by light have zero temperature, $T_e = 0$, so that *cold* electrons occupy the Fermi sphere in *k*-space with the radius $k_F = \sqrt{2m_i\varepsilon_F}/\hbar$, see Fig.5a. Then the *excited* ("hot") electrons occupy a spherical layer above the Fermi sphere in *k*-space:

$$k_F = \sqrt{2m_i\varepsilon_F}/\hbar < k < k_{\hbar\omega} = \sqrt{2m_i(\varepsilon_F + \hbar\omega)}/\hbar \quad (27)$$

(see Fig. 5). If the hot electron, in its final state after photon absorption, has the energy $E_f = \hbar^2 k_f^2/(2m_i)$ ($\mathbf{k}_f$ is the wave vector of the hot electron) larger than the height $U_b$ of potential barrier, i.e.

$$E_f = \hbar^2 k_f^2/(2m_i) > U_b \equiv \hbar^2 k_{bar}^2/(2m_i), \quad (28)$$

(see Fig.5a), then it has a chance to leave the nanoparticle. The ratio of the emission rate $R_{em}^{volume}$ of hot electrons from the nanoparticle to the excitation rate of hot electrons inside the nanoparticle, which simply equals to the photon absorption rate $R_{abs}^{volume}$, is, by definition, the *internal quantum efficiency* $\eta_i$ of the volume photoelectric effect [7],

$$\eta_i = R_{em}^{volume}/R_{abs}^{volume}, \quad (29)$$

Below we calculate $\eta_i$ for a spherical nanoparticle.

We will assume in calculations that the distribution of hot electron in the layer in *k*-space is *uniform* and *isotropic*. This assumption is key point in Fowler's statistical theory of photoelectric effect from metals [55], and we are adopting this assumption here (see also modeling in [7]). Then, the density of electron excitation rate in *k*-space, in units of $\left[m^3/(s\cdot m^3) = 1/s\right]$, is

$$r_{exc}^{(k)} = \frac{r_{abs}^{vol}}{V_{layer}^{(k)}}, \quad (30)$$

where

$$V_{layer}^{(k)} = \frac{4\pi}{3}\left(k_{\hbar\omega}^3 - k_F^3\right), \qquad (31)$$

is the volume occupied in $k$-space by hot electrons.

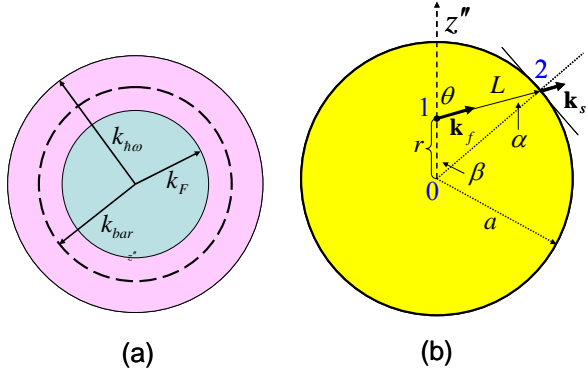

Fig. 5. (a) Distribution of electrons in k-space. Blue color shows the volume in *k*-space occupied by unexcited (cold) electrons; pink color illustrates excited (hot) electrons after photon absorption. Hot electrons outside dash line can be emitted from nanoparticle. (b) Illustration of propagation of hot electron in spherical nanoparticle of the radius $a$. Hot electron is generated at the distance $r$ from the center 0 with the wave vector $\mathbf{k}_f$ with the angle $\theta$ to the axis $z''$. In absence of collisions, the electron moves along straight line parallel to $\mathbf{k}_f$ and collides with the nanoparticle surface at the point 2. $L$ is the length of the electron path before its collision with the nanoparticle boundary. $\alpha$ is the incidence angle of the electron to the surface. $\mathbf{k}_s$ is the wave vector of the electron as it collides with the surface.

If $p_{em}(\mathbf{r},\mathbf{k}_f)$ is the probability for the hot electron (with the wave vector $\mathbf{k}_f$, generated in the nanosphere at the position $\mathbf{r}$) to be emitted from the nanoparticle, the photoelectron emission rate from the nanosphere is

$$R_{em}^{volume} = \int_{volume} d\mathbf{r} \int_{layer} d\mathbf{k}_f\, r_{exc}^{(k)} \cdot p_{em}(\mathbf{r},\mathbf{k}_f) = R_{abs}^{volume} \cdot \eta_i, \quad (32)$$

where $\eta_i$ is the internal quantum efficiency,

$$\eta_i = \frac{1}{V_{layer}^{(k)}}\frac{1}{V_{nano}}\int_{volume} d\mathbf{r} \int_{layer} d\mathbf{k}_f\, p_{em}(\mathbf{r},\mathbf{k}_f), \qquad (33)$$

In our derivation of (32)-(33), we used Eqs. (26) and (30). While the electron moves towards the boundary of the nanoparticle, it can experience elastic and inelastic collisions with phonons and cold electrons, and finding $p_{em}(\mathbf{r},\mathbf{k}_f)$ is quite a complicated problem of physical kinetics. References to papers, where various approaches to solve this problem were employed, can be found in [7, 38, 56-57]. In this paper, we are using the simple model, presented in [56], in order to find the probability $p_{em}(\mathbf{r},\mathbf{k}_f)$ and to calculate the efficiency $\eta_i$. Let us consider at first the case when electron moves to the boundary freely, i.e. without collisions. Fig. 5b illustrates this. A hot electron is generated with the wave vector $\mathbf{k}_f$ at the point 1 in the sphere at the distance $r < a$ from its centre (0). The vector $\mathbf{k}_f$ is directed at the angle $\theta$ to the axis $z''$, which goes from the center 0 and passes through the point 1. Because we assume "collisionless" motion of the electron, it moves along a straight line parallel to the vector $\mathbf{k}_f$. The electron collides with the spherical nanoparticle boundary at the point 2 under the incidence angle $\alpha$ – see Fig.5b.

In general, if the electron arrives at the boundary with the wave vector $\mathbf{k}_s$, the probability $p_{em}(\mathbf{r},\mathbf{k}_f)$ is equal to the quantum mechanical probability (transmission) $t_{bar}(\mathbf{k}_s)$ for this electron to overcome the potential barrier at the boundary between metal and surrounding medium, $p_{em}(\mathbf{r},\mathbf{k}_f) \equiv t_{bar}(\mathbf{k}_s)$. The transmission $t_{bar}(\mathbf{k}_s)$ is a function of the component $k_s^{(n)}$ of the vector $\mathbf{k}_s$ normal to the nanoparticle surface at the point of collision between the electron and the surface: $t_{bar}(\mathbf{k}_s) \equiv t_{bar}(k_s^{(n)})$. It is well-known that the transmission $t_{bar}(k_s^{(n)})$ depends strongly on the shape of the potential at the nanoparticle boundary.

In the "collisionless" case of electron motion which we consider at first, the wave vector $\mathbf{k}_s$ is equal to $\mathbf{k}_f$. Therefore, $k_s^{(n)} = k_f \cos\alpha$, and

$$p_{em}^{no\,coll}(\mathbf{r},\mathbf{k}_f) = t_{bar}(k_f \cos\alpha), \qquad (34)$$

In this case, the six-fold integral in (33) can be easily converted to a triple integral

$$\eta_i = \frac{1}{V_{layer}^{(k)}}\frac{1}{V_{nano}}\int_0^a dr\, 4\pi r^2 \int_{k_F}^{k_{\hbar\omega}} dk_f \int_0^{\pi} d\theta\cdot\sin\theta\, t_{bar}(k_f\cdot\cos\alpha), \qquad (35)$$

where the incidence angle $\alpha$ is related with the angle $\theta$ by the theorem of sinuses:

$$\alpha = \arccos(r\sin\theta/a), \qquad (36)$$

The integral in (35) can be calculated analytically for some shapes of the potential – see below.

Now, having considered the simplest collision-free case, we move on to consider a more realistic and more complicated case when the electron can experience collisions during its motion to the surface and is therefore cooled. It is well-known that the dominating mechanism for cooling of hot electrons is their collisions with cold electrons [58-60]. In fact, *just one* collision of a hot electron with a cold electrons renders the hot electron unable to overcome the potential barrier between metal and surrounding medium. Therefore, in order to calculate the photoelectron emission probability $p_{em}(\mathbf{r},\mathbf{k}_f)$ in the case when collisions of hot electron are possible, we can simply multiply the "collisionless" probability $p_{em}^{no\,coll}(\mathbf{r},\mathbf{k}_f)$ [see Eq.(34)] by the probability $P_t(\mathbf{r},\mathbf{k}_f)$ that the hot electron reaches the surface *without* collisions:

$$p_{em}(\mathbf{r},\mathbf{k}_f) = P_t(\mathbf{r},\mathbf{k}_f)\cdot t_{bar}(k_f\cos\alpha), \qquad (37)$$

Following [56], the probability $P_t(\mathbf{r}, \mathbf{k}_f)$ can be written as

$$P_t(\mathbf{r}, \mathbf{k}_f) = \exp\left[-L(r,\theta)/l_e(E_f)\right], \quad (38)$$

where

$$L(r,\theta) = \sqrt{a^2 - r^2 \cdot \sin^2\theta} - r\cos\theta, \quad (39)$$

is the distance between the generation point 1 and the point 2 where electron collides with the nanoparticle surface, as it moves in metal without collisions (see Fig.5b); the mean free path $l_e(E_f)$, generally speaking, depends on the hot electron energy $E_f$ [56-60]. Correspondingly, from (33) with taking into account (37), we have

$$\eta_i = \frac{1}{V_{layer}^{(k)}} \frac{1}{V_{nano}} \int_0^a dr\, 4\pi r^2 \int_{k_F}^{k_{\hbar\omega}} dk_f \int_0^\pi d\theta \cdot \sin\theta \cdot$$
$$\cdot 2\pi k_f^2 \exp\left[-L(r,\theta)/l_e(E_f)\right] \cdot t_{bar}(k_f \cdot \cos\alpha) \quad (40)$$

Obviously, if $l_e = \infty$, Eq.(40) coincides with Eq.(35).

Using (2), (1), (32), (25), we can finally express the photoelectron emission cross-section for the volume photoelectric effect through the internal quantum efficiency $\eta_i$ as

$$\sigma_{em}^{volume} = \frac{8\pi^2}{3} \frac{a}{\lambda_o} \frac{\varepsilon_i''}{\sqrt{\varepsilon_e}} \cdot |F|^2 \eta_i \cdot a^2, \quad (41)$$

where $\lambda_o = 2\pi c/\omega$ is the light wave length in vacuum.

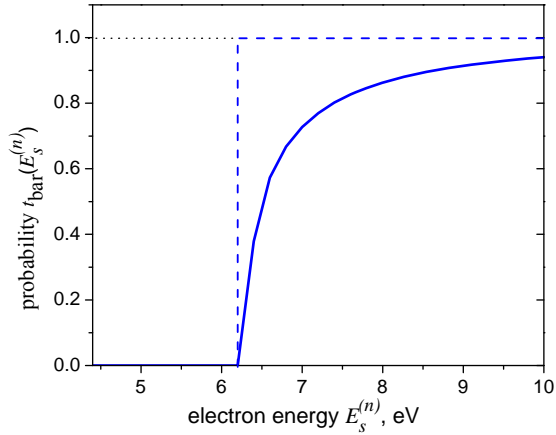

Fig. 6. The probability (transmission) $t_{bar}$ for an electron to leave metal for the step barrier potential with $U_b = 6.31$ eV as function of the electron energy $E_s^{(n)}$, i.e. $t_{bar} = t_{bar}(E_s^{(n)})$ – blue curve. Dashed blue curve is the model function for the probability given by Eq.(44) ("model 0-1").

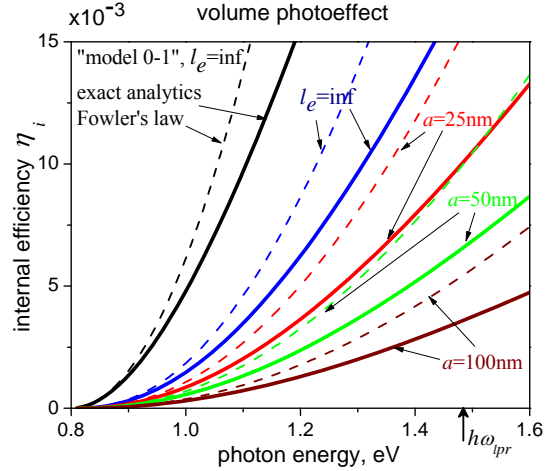

Fig.7. The spectral dependence of the internal quantum efficiency $\eta_i(\hbar\omega)$ for volume photoelectric effect for nanoparticles with different radiuses $a$ and different potential shapes at the nanoparticle surface. Red, green and brown solid curves are calculated for $a = 25, 50$ and $100$ nm, respectively, by numerical integration in Eq.(40) for the step potential [see Eqs.(42)-(43)] and using the mean free path $l_e = 41$ nm, while dash curves are obtained with the approximation (45). Blue and black curves are obtained for $l_e = \infty$ (i.e. for collisionless electron propagation in nanoparticle) for the step potential and the 'model 0-1' [see Eq.(44)], respectively. Blue and black solid lines are exact results given by Eq.(40) and (47), respectively, while blue and black dash curves are obtained with approximations (45) and (48), respectively. $\hbar\omega_{lpr}$ is the photon energy when the plasmonic resonance occurs in the nanoparticle.

As mention above, the probability $t_{bar}(k_s^{(n)})$ depends on the shape of the potential $U(z)$ at the nanoparticle surface. Since we aim to compare surface and volume photoelectric effects, and the surface effect have been calculated above for the step potential (see Fig. 3), below we calculate the internal quantum efficiency $\eta_i$ firstly for this shape of potential. In this case, the probability $t_{bar}$ is

$$t_{bar} = \frac{4\,\text{Re}[s]}{|1+s|^2}, \quad (42)$$

where

$$s = \sqrt{1 - \frac{2mU_b}{\hbar^2\left(k_s^{(n)}\right)^2}} \equiv \sqrt{1 - \frac{U_b}{E_s^{(n)}}} \quad (43)$$

Fig. 6 illustrates the dependence of the transmission $t_{bar}$ on the energy $E_s^{(n)} = \hbar^2\left(k_s^{(n)}\right)^2/(2m)$. When $E_s^{(n)} > U_b$, the transmission at first increases sharply and then tends to 1 gradually. On the other hand, many papers (see for instance [7]) assume a simpler *model* dependence:

$$t_{bar}\left(E_s^{(n)}\right) = \begin{cases} 0, & E_s^{(n)} < U_b \\ 1, & E_s^{(n)} > U_b \end{cases}, \quad (44)$$

(see dashed line in Fig.5a), which is a suitable model for smoothly changing potential (see dashed green line in Fig. 3) rather than for sharply changing one. Below we refer to the model (44) as "model 0-1".

Thick solid curves in Fig. 7 show the spectral dependences of the internal quantum efficiency $\eta_i(\hbar\omega)$, obtained by numerical integration in Eq.(40) with the barrier transmission $t_{bar}(E_s^{(n)})$ for the step potential [see Eqs.(42)-(43)] for various radiuses $a$ of the spherical nanoparticle: red line is for $a = 25$ nm, green line is $a = 50$ nm, brown line is $a = 100$ nm. In calculations we used the same material parameters for nanoparticle and the barrier as in Fig. 4. For gold, in concerned range of the hot electron energies $E_f$ the mean free path $l_e(E_f)$ changes very weakly, from ~42 to ~40 nm [60]. Therefore, we have considered the value $l_e$ as constant, and used $l_e = 41$ nm. For comparison, thick blue line shows the efficiency $\eta_i$ for the mean free path $l_e = \infty$, i.e. for the "collisionless" propagation of hot electrons in nanoparticle. One sees that electron collisions decreases the efficiency $\eta_i$ by several times for the shown values of radius $a$.

Close to the photoeffect threshold ($\hbar\omega \to W_b$), one can do the integration in Eq.(40) analytically, and obtain the approximate formula for the internal efficiency for $\eta_i$:

$$\eta_i = F_{st}(a, l_e) \times \frac{12}{5} \frac{(\varepsilon_F + W_b)^{3/2}}{(\varepsilon_F + \hbar\omega)^{3/2} - \varepsilon_F^{3/2}} \cdot \left(\frac{\hbar\omega - W_b}{\varepsilon_F + W_b}\right)^{5/2}, \quad (45)$$

where

$$F_{st}(a, l_e) = \frac{l_e}{2a}\left[1 - \exp\left(-\frac{2a}{l_e}\right)\right], \quad (46)$$

is the *structural* function describing the dependence of the efficiency $\eta_i$ on the nanoparticle size $a$. The dependence (45) is illustrated in Fig. 7 by dashed blue, red, green and brown curves. Parameters used in calculation on these curves are the same as in calculation of solid curves of the same color. Thus, $\eta_i \propto (\hbar\omega - W_b)^{5/2}$ close to the threshold, and this behaviour coincides with behaviour of the parameter $\eta_o$ in the surface effect [see Eqs.(21)-(22)], calculated also for the step potential.

As mention above the model dependence (44) for the transmission $t_{bar}(E_s^{(n)})$ is used often. With this $t_{bar}(E_s^{(n)})$, the integral in Eq.(40) can be calculated analytically for arbitrary excess $(\hbar\omega - W_b)$ over the photoeffect threshold, and is expressed through the exponential integral function $\text{Ei}(z)$ − we do not give the result due to its bulkiness. But for $l_e = \infty$ ("collisionless"

electron propagation in nanoparticle) we have simple approximate formula

$$\eta_i = \frac{(\varepsilon_F + W_b)^{3/2}}{(\varepsilon_F + \hbar\omega)^{3/2} - \varepsilon_F^{3/2}} \cdot \left[\left(\frac{\varepsilon_F + \hbar\omega}{\varepsilon_F + W_b}\right)^{3/2} - \frac{3}{2}\log\frac{\varepsilon_F + \hbar\omega}{\varepsilon_F + W_b} - 1\right], \quad (47)$$

On other hand, for arbitrary $l_e$, but near to threshold we have the approximation

$$\eta_i \approx F_{st}(a, l_e) \times \frac{9}{8} \cdot \frac{(\varepsilon_F + W_b)^{3/2}}{(\varepsilon_F + W_b)^{3/2} - \varepsilon_F^{3/2}} \cdot \left(\frac{\hbar\omega - W_b}{\varepsilon_F + W_b}\right)^2, \quad (48)$$

These analytic results (47) and (48) for $l_e = \infty$ are illustrated in Fig.7 by black solid and dash curves, respectively. Obviously, near to the photoeffect threshold Eq.(48) yields the parabolic Fowler's law: $\eta_i \propto (\hbar\omega - W_b)^2$. Formulas (47) and (48) with $l_e = \infty$ are *spherical* analogs of the corresponding formulas for the internal quantum efficiency for bulk photoelectron emission from *thin-films* which can be found in [7] when electron collisions are neglected.

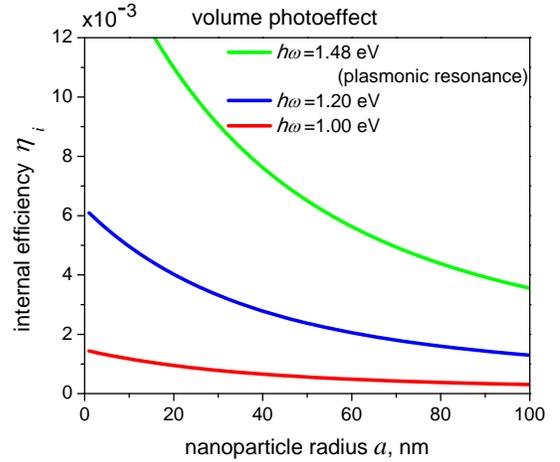

Fig. 8. Dependence of the internal quantum efficiency $\eta_i$ on the nanoparticle radius $a$ for three different photon energies: $\hbar\omega = 1.0$, 1.2 and 1.48 eV. The mean free path $l_e = 41$ nm. For $\hbar\omega = 1.48$ eV, the plasmonic resonance takes place in the gold spherical nanoantenna.

Fig. 8 shows the dependences of the internal efficiency $\eta_i$ on the nanoparticle radius $a$ for different photon energies: $\hbar\omega = 1.0$ and 1.2 eV, and at the plasmonic resonance $\hbar\omega = \hbar\omega_{lpr} = 1.48$ eV. The behavior of the curves is described, at least qualitatively, by the structural function (46), and for large $a$, the efficiency follows $\eta_i \propto 1/a$. This last circumstance originates from the fact that within the quasistatic approximation the photon absorption rate $R_{abs}^{volume}$ is proportional to the nanoparticle volume $V_{nano}$ [see Eq.(25)] while for large radiuses $a \gg l_e$, obviously, the photoelectron emission rate $R_{em}^{volume}$ is proportional to the nanoparticle surface $A_{nano}$. Correspondingly, $\eta_i \propto A_{nano}/V_{nano} \propto 1/a$. Thus, the behavior $\eta_i \propto 1/a$ is a consequence of the fact that the "surface to volume" ratio decreases with increasing nanoparticle size $a$. On the other hand, for smaller radiuses $a$ when $R_{em}^{volume} \propto V_{nano}$, the efficiency $\eta_i$ tends to its values in the "collisionless" case – see Eq.(35).

## III. Comparison of surface and volume mechanisms and Discussion

Surface and volume mechanism of photoelectron emission from nanoparticles can be compared by considering the ratio of their photoelectron emission cross-sections:

$$K_{v-s} = \frac{\sigma_{em}^{volume}}{\sigma_{em}^{surfave}} = \frac{2\pi a}{\lambda_o} \cdot \frac{\varepsilon_i'' \eta_i}{\eta_o}, \quad (49)$$

where we used Eqs. (23) and (41). Fig. 9 shows the spectral dependences of the ratio $K_{v-s}(\hbar\omega)$ for different nanoparticle radiuses: $a = 100, 50$ and $25$ nm for blue, red and brown curves, respectively. Dashed curves are obtained when electron collisions are neglected ($l_e = \infty$), and solid lines are calculated with $l_e = 41$ nm.

The dashed curves show that if cooling of hot electrons is absent, the volume photoeffect from nanoparticles can predominate, at least for larger nanoparticles far from the photoeffect threshold. However, if we take into account hot electron cooling during their propagation in nanoparticle (see solid curves in Fig. 9), we obtain that the surface mechanism is stronger than the volume one almost throughout the considered spectrum range, and only at $\hbar\omega > 1.5$ eV the ratio $K_{v-s}$ becomes larger than one. One should note that in considered case of gold nanosphere buried into a medium with $\varepsilon_e = 13$ the localized plasmonic resonance occurs at $\hbar\omega \approx \hbar\omega_{lpr} = 1.48$ eV (shown in Fig. 9). Thus, at the plasmonic resonance frequency in given structure the surface and volume mechanism give comparable contributions into the photoelectron emission rate.

One should stress that the results in Fig. 9, and in particular, the dashed curves for "collisionless" electron propagation in nanoparticle are obtained in the quasistatic approximation [52], when the electromagnetic field efficiently penetrates into nanoparticle. In this case, the volume photoelectron emission rate increases proportionally to the nanoparticle volume if we neglect electron cooling. On the other hand, the surface photoelectron emission rate increases proportionally to the nanoparticle surface area. Correspondingly, if we neglect the electron collisions, the growing role of the volume mechanism with increasing nanoparticle size can be attributed to the quasistatic approximation. In fact, this approximation works only for relatively small plasmonic nanoparticles, and already for $a = 100$ nm the field penetrates into nanoparticle volume only partially [52]. Thus, the model used here can seriously overestimate the volume mechanism for large nanoparticles if we neglect the hot electron cooling, as shown by dashed curves in Fig. 9.

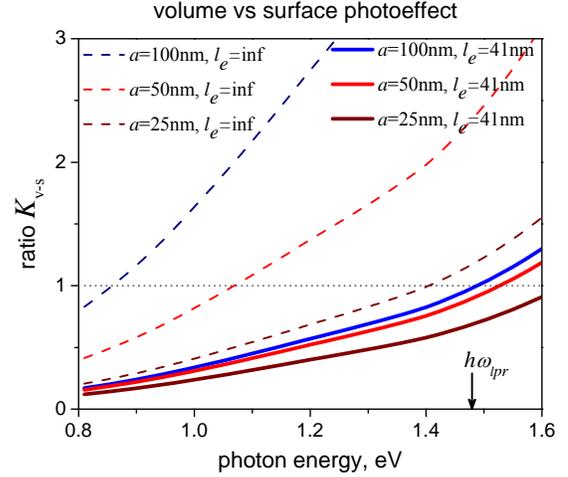

Fig. 9. Spectral dependence the ratio $K_{v-s} = \sigma_{em}^{volume}/\sigma_{em}^{surfave}$ for various values of the nanoparticle radius $a$ and the mean free path $l_e$. For the plasmonic resonance, $\hbar\omega \approx \hbar\omega_{lpr} = 1.48$ eV.

Fig.10 shows the ratio $K_{v-s}$ as a function of the nanoparticle radius $a$ for different photon energies: $\hbar\omega = 1.0, 1.2$ eV and in the case of plasmonic resonance $\hbar\omega = \hbar\omega_{lpr} = 1.48$ eV. One sees that the ratio $K_{v-s}$ tends to zero with the decreasing radius $a$. Such behavior again results from the geometric relation of the ratio "surface to volume" at small $a$. Namely, for $a \ll l_e$ the volume photoelectron emission rate is $R_{em}^{volume} \propto V_{nano}$, while the surface photoelectron emission rate is $R_{em}^{surface} \propto A_{nano}$. Therefore, $K_{v-s} = R_{em}^{volume}/R_{em}^{surfave} \propto V_{nano}/A_{nano} \propto a$. On the other hand, one sees from Fig. 9 that for larger values of the radius $a$ ($a \gg l_e$) when the hot electrons generated far from the surface can not reach it due to cooling by collisions, both the rates $R_{em}^{volume}$ and $R_{em}^{surface}$ become proportional to the nanoparticle surface area $A_{nano}$, and correspondingly, the ratio $K_{v-s}$ does not change with the radius $a$ any more.

Now we would like to discuss the assumptions made in the calculation of the volume mechanism in more detail. The first one is the quasistatic approximation, which leads to overestimation of the volume effect for larger nanoparticles as discussed above. Another assumption concerns the distribution of hot electrons in $k$-space – see Eqs.(30)-(31). In fact, we ascribe the light absorption, which is defined by the experimentally measured value of the imaginary part $\varepsilon_i''(\omega)$ of the metal relative permittivity, uniformly to *all* possible optical transitions from the states under the Fermi level; as a result, the hot electrons populate the volume in $k$-space, given by Eq. (27), equally. This means that the hot electron distribution on the energy $E_f$ is proportional to $\sqrt{E_f}$. On other hand, Chen and Bates [56] consider that the distribution is proportional to "joint density of states" which is $\propto \sqrt{E_f(E_f - \hbar\omega)}$. Our analysis shows that use of the latter distribution instead of the Fowler's one can increase the volume effect but only slightly – by 2-7%.

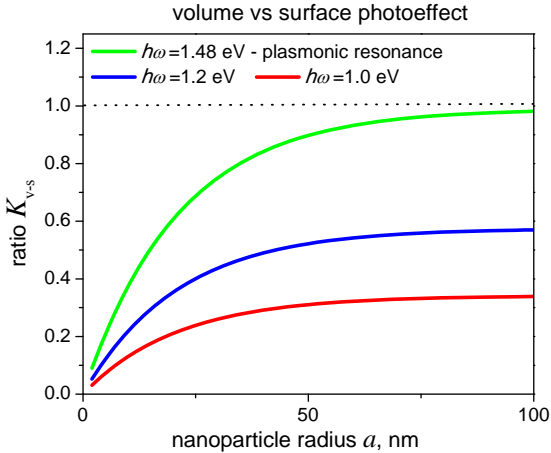

Fig. 10. Dependence the ratio $K_{v-s} = \sigma_{em}^{volume}/\sigma_{em}^{surface}$ on the nanoparticle radius $a$ for three different photon energies: $\hbar\omega = 1.0$, 1.2 and 1.48 eV. The mean free path $l_e = 41$ nm.

Furthermore, it is well-known that optical transitions from deeper states under Fermi level can contribute to the optical absorption more strongly than the transitions from states that lie closer to the Fermi level [38, 61]. It is clear that the transitions from deeper levels populate states over Fermi level with *lower* energies than it is done by transitions from less deep levels. Hence, lower-energy states above the Fermi level are populated with hot electrons more strongly than the higher-energy levels; it is only from these latter levels that the electron emission from metal is possible in principle. In others words, transitions from deeper levels can contribute strongly to light absorption in metals but not to the photoelectron emission since such transitions do not generate hot electrons of sufficient energy so as to overcome the barrier. Thus, the assumption on homogeneous population (30) of the layer (27) (the assumption by Fowler) can lead, generally speaking, to an overestimation of the internal quantum efficiency $\eta_i$ for the volume mechanism. This can particularly concern the interband transitions in metal [61]. Note that the thought that the optical transitions of electrons in the bulk can determine the optical absorption of metal, but are unable to provide photoelectron emission, was one of the basic ideas of the paper by Tamm and Schubin [36]. However, these strong transitions from deeper energy level in gold appear to become substantial only at photon energies higher than ~1.5-2eV [54, 61], so that for the photon energy range under consideration (i.e. $\hbar\omega < 1.6\,\text{eV}$) the assumption of uniform population of states by hot electrons looks rather reasonable.

On the other hand, the model (37)-(39) for hot electron cooling by their collusions with cold electrons can underestimate the electron cooling rate. Indeed, the model neglects, in some sense, the role of the elastic collisions of hot electron. Such collisions do not affect the hot electron energy directly, but can make the electron path in the nanoparticle longer than the straight line (39), which can lead to stronger electron cooling rate.

Nevertheless, despite the limitations brought about by the above assumptions, we can suggest that it appears to be unlikely that more precise models would significantly alter the conclusion that we draw from the calculations in this work, namely, that the cooling of hot electrons leads to a serious decrease of the photoelectron emission rate in the volume mechanism. Together with the fact that the surface mechanism of the photoelectric effect does not suffer from electron collisions in the bulk at all, this lets us conclude that the surface mechanism prevails over the volume one in plasmonic nanoantennas for the majority of considered conditions.

Let us finally recall that our the case when the surface and volume mechanisms appear to be comparable to each other, i.e., under the plasmonic resonance condition (see Fig. 9 and 10), is obtained for *spherical* nanoparticles. It is well known that the plasmonic resonance in *oblate spheroids*, which are good models for nanoparticles in many experimental configurations, is red-shifted if incident light is polarized along the longer axis [6, 52]. In accordance with Fig. 9, one could assume that in such nanoantennas the surface photoelectric effect can prevail also in the case of plasmonic resonance. Numerical calculations of photoelectron emission from nanoparticles with shapes more complicated than spherical, as well as from ensembles of nanoparticles, are subject of planned forthcoming studies.

IV. Conclusion

We have carried out calculations of photoelectron emission from plasmonic nanoparticles into surrounding semiconductor matrix for the IR range of photon energies for two mechanisms of the effect – surface and volume ones – and showed that the surface photoeffect prevails over the volume one, confirming the initial conclusion by Tamm and Schubin in 1931 in their pioneering development of quantum-mechanical theory of photoeffect from metal. From our calculations, this predominance of the surface effect is a result of effective cooling of generated hot carriers during their propagation from the inside of the nanoparticle to its surface in the necessary for the scenario of the volume mechanism. Calculations have been done for the step potential at the nanoparticle surface, and a simple model for the hot electron cooling has been used. Nevertheless, it is unlikely that these model limitations would change the conclusion. To our knowledge such comparison of two mechanisms of the photoeffect from plasmonic nanoparticles was done for the first time.

We also stress the effect of the discontinuity of the dielectric permittivity at the nanoparticle boundary in the surface mechanism, which leads to substantial (by ~5 times) increase in photoelectron emission rate from nanoparticle compared to the case when such discontinuity is absent.

Confidence in the predominance of the surface mechanism can be useful in the design of devices based on photoelectric effect and on the use of hot electrons from plasmonic nanoantennas. In particular, it is clear that the angular pattern of photoelectrons can be different for the surface and volume photoeffect. For instance, for spherical nanoparticles, in the case of volume mechanism electrons are emitted into all directions equally, while in the case of surface one electrons are emitted mainly into direction parallel to the electric field polarization of incident light because the photoelectron emission rate in the surface mechanism is proportional to the square of the electric field component normal to the nanoparticle surface. Correspondingly, parts of nanoantenna surface where the normal component is maximal must have good contact to surrounding matrix to which the electron emission can occur. It appears that this circumstance played a key role in [12] where substantial increase of the responsibility of the device was reached with proper embedding of nanoantennas into the semiconductor substrate.

## Acknowledgments

A.V.U. and I.E.P. acknowledge financial support from the Russian Foundation for Basic Research (Project No. 14-02-00125) and the Russian MSE State Contract N14.527.11.0002 and support from the CASE project (Denmark). V.E.B. acknowledges financial support from SPIE Optics and Photonics Education Scholarship, as well as Otto Mønsteds and Kaj og Hermilla Ostenfeld foundations. S.V.Z. acknowledges financial support from the People Programme (Marie Curie Actions) of the European Union's 7th Framework Programme FP7-PEOPLE-2011-IIF under REA grant agreement No. 302009 (Project HyPHONE).

## References

[1] H. Chalabi and M. L. Brongersma, Nature Nanotechnology, **8**, 229 (2013)
[2] Y. Nishijima, K. Ueno, Y. Yokota, Kei Murakoshi, and H. Misawa, J. Phys. Chem. Lett., **1**, 2031 (2010)
[3] M. W. Knight, H. Sobhani, P. Nordlander, and Naomi J. Halas, Science, **332**, 702 (2011).
[4] Y. Takahashi and T. Tatsuma, Appl. Phys. Lett., **99**, 182110 (2011)
[5] Y. K. Lee, Chan Ho Jung, J. Park, H. Seo, G. A. Somorjai, and J. Y. Park, Nano Lett., **11**, 4251 (2011).
[6] I. E. Protsenko and A. V. Uskov, Phys. Usp., **55**, 508 (2012).
[7] C. Scales and P. Berini, IEEE J. Quantum Electron., **46**, 633 (2010).
[8] A. Akbari, and P. Berini, Appl. Phys. Lett., **95**, 021104 (2009).
[9] P. Berini, Laser Photonics Rev., pp. 1-24, 14 June 2013, DOI: 10.1002/lpor.201300019
[10] I. Goykhman, B. Desiatov, J. Khurgin, J. Shappir, and U. Levy, Nano Lett., **11**, 2219 (2011)
[11] A. Novitsky, A. V. Uskov, C. Gritti, I. E. Protsenko, B. E. Kardynał, and A. V. Lavrinenko, Prog. Photovolt.: Res. Appl., DOI: 10.1002/pip.2278 (2012).
[12] M. W. Knight, Y. Wang, A. S. Urban, A. Sobhani, B. Y. Zheng, P. Nordlander, and N. J. Halas, Nano Lett. **13**, 1687 (2013).
[13] A. Sobhani, M. W. Knight, Y. Wang, B. Zheng, N. S. King, L. V. Brown, Z. Fang, P. Nordlander, and N. J. Halas, Nature Comm., **4**, 1643 (2013).
[14] S. V. Zhukovsky, V. E. Babicheva, A. V. Uskov, I. E. Protsenko, and A. V. Lavrinenko, Plasmonics, Oct. 2013, DOI: 10.1007/s11468-013-9621-z.
[15] E. Moulin, P. Luo, B. Pieters, J. Sukmanowski, J. Kirchhoff, W. Reetz, T. Müller, R. Carius, F.-X. Royer, and H. Stiebig, Appl. Phys. Lett., **95**, 033505 (2009).
[16] E. A. Moulin, U. W. Paetzold, B. E. Pieters, W. Reetz, and R. Carius, J. Appl. Phys. **113**, 144501 (2013).
[17] T. P. White and K. R. Catchpole, Appl. Phys. Lett., **101**, 073905 (2012).
[18] F. Wang and N. A. Melosh, Nano Lett., **11**, 5426 (2011).
[19] A. K. Pradhan, T. Holloway, R. Mundle, H. Dondapati, and M. Bahoura, Appl. Phys. Lett., **100**, 061127 (2012).
[20] F. P. Garcia de Arquer, A. Mihi, D. Kufer, and G. Konstantatos, ACS Nano, **7**, 3581, (2013).
[21] F. B. Atar, E. Battal, L.E. Aygun, B. Daglar, M. Bayindir, and A. K. Okyay, Opt. Express, **21**, 7196 (2013).
[22] M. Grätzel, Nature, **414**, 338 (2001).
[23] E. W. McFarland and J. Tang, Nature, **421**, 616 (2003).
[24] J. R. Renzas and G. A. Somorjai, J. Phys. Chem. C, **114**, 17660 (2010).
[25] Y. Li, and G. A. Somorjai, Nano Lett., **10**, 2289 (2010).
[26] I. Thomann, B. A. Pinaud, Z. Chen, B. M. Clemens, Th. F. Jaramillo, and M. L. Brongersma, Nano Lett., **11**, 3440 (2011).
[27] S. Mukherjee, F. Libisch, Nicolas Large, O. Neumann, L. V. Brown, J. Cheng, J. B. Lassiter, E. A. Carter, P. Nordlander, and N. J. Halas, Nano Lett., **13**, 240 (2012).
[28] S. C. Warren and E. Thimsen, Energy Environ. Sci., **5**, 5133 (2012).
[29] S. Mubeen, J. Lee, N. Singh, S. Krämer, G. D. Stucky and M. Moskovits, Nature Nanotechnology, **8**, 247 (2013).
[30] M. Xiao, R. Jiang, F. Wang, C. Fang, J. Wang and J. C. Yu, J. Mater. Chem. A, **1**, 5790 (2013).

[31] D. Conklin, S. Nanayakkara, T.-H. Park, M. F. Lagadec, J. T. Stecher, X C., M. J. Therien, and D. A. Bonnell, ACS Nano, **7**, 4479 (2013).
[32] J. W. Schwede, I. Bargatin, D. C. Riley, B. E. Hardin S. J. Rosenthal, Y. Sun, F. Schmitt, P. Pianetta, R. Howe, Z.-X. Shen, and N. A. Melosh, Nature Mater., **9,** 762 (2010).
[33] C. Ropers, T. Elsaesser, G. Giulio Cerullo, M. Zavelani-Rossi, and C. Lienau, New J. Phys. **9,** 397 (2007)
[34] F.-J. Meyer zu Heringdorf, L.I. Chelaru, S. Möllenbeck, D. Thien and M. Horn-von Hoegen, Surface Science, **601,** 4700 (2007).
[35] M. I. Stockman, M. F. Kling, U. Kleineberg, and F. Krauz, Nature Photonics, **1,** 539 (2007).
[36] I. Tamm and S. Schubin, Zeischrift für Physik **68**, 97 (1931).
[37] U. Kreibig and M. Vollmer, "Optical Properties of Metal Clusters", Springer-Verlag Berlin Heidelberg (1995)
[38] N. V. Smith, CRC Critical Reviews in Solid State Sciences, **2**, 45 (1971).
[39] C. Kittel, Quantum Theory of Solids, 2$^{nd}$ Rev. Ed., John Willey & Sons, 1989.
[40] J.M. Ziman, Principles of the Theory of Solids, 2$^{nd}$ Ed., Cambridge University Press, 1972.
[41] K. Mitchell, Proc. of the Royal Society of London. Series A, **146**, 442 (1934).
[42] R. E. B. Makinson, Phys. Rev., **75**, 1908 (1949).
[43] A. M. Brodsky and Yu. Ya. Gurevich, Sov. Phys. JETP, **27**, 114 (1968).
[44] A. M. Brodsky, Yu. Ya. Gurevich, and V.G. Levich, Physica Status Solidi, **40**, 139 (1970).
[45] A. M. Brodsky and Yu. Ya. Gurevich, Theory of Electron Emission from Metals, Nauka: Moscow, 1973.
[46] J. G. Endriz and W. E. Spicer, Phys. Rev. Lett., **27**, 570 (1971).
[47] J. G. Endriz, "Photoemission studies of surface-plasmon oscillations on controlled-roughness aluminum films", Ph.D. thesis, Stanford University, 1970.
[48] J. G. Endriz, Phys. Rev. B, **7**, 3464(1973)
[49] A. A. Maradudin and D. L. Mills, Phys. Rev. B, **11**, 1392 (1975).
[50] H. Petersen, Z. Physik B, **31**, 171 (1978).
[51] A. O. Govorov, H. Zhang, and Yu. K. Gun'ko, J. Phys. Chem. C, **117**, 16616 (2013).
[52] S. A. Maier, Plasmonics: Fundamentals and Applications. Springer Science+Business Media LLC, 2007.
[53] V. L. Ginzburg and I. M. Frank, JETP, **16**, 15 (1946).
[54] M. J. Weber and Marvin J., Handbook of optical materials, CRC Press LLC, 2003
[55] R. H. Fowler, Phys. Rev., **38,** 45 (1931)
[56] Q. Y. Chen and C. W. Bates, Phys. Rev. Lett., **57**, 2737 (1986).
[57] C. N. Berglund and W. E. Spicer, Phys. Rev. , **136**, 1030 (1964).
[58] R. Stuart, F. Wooten, and W. E. Spicer, Phys. Rev., **135**, 495 (1964).
[59] J. J. Quinn, Phys. Rev., **126**, 1453 (1962).
[60] K. W. Frese and C. Chen, J. Electrochem. Soc., **139**, 3234 (1992).
[61] G. V. Hartland, Chem. Rev., **111**, 3858 (2011).